\documentclass[Times,4pt,aps,pra,twocolumn,showpacs,amsmath,amssymb,floatfix,footinbib,superscriptaddress]{revtex4-1}
\usepackage{amsmath,natbib,graphicx,subfigure,amssymb,graphics,amsmath,mathrsfs,CJK,color}
\usepackage{multirow,fancyhdr,color,bm,tabularx,psfrag,geometry,dcolumn,datetime}
\usepackage[ colorlinks=true,linkcolor=blue,urlcolor=blue,citecolor=blue]{hyperref}
\usepackage[mathlines]{lineno}
\def\footnoterule{\kern -1mm \hrule width 5.8cm \kern 2.2mm}
\usepackage{tikz,xcolor,hyperref}
\definecolor{lime}{HTML}{A6CE39}
\DeclareRobustCommand{\orcidicon}{%
    \begin{tikzpicture}
    \draw[lime, fill=lime] (0,0)
    circle [radius=0.16]
    node[white] {{\fontfamily{qag}\selectfont \tiny ID}};\draw[white, fill=white] (-0.0625,0.095)
    circle [radius=0.007];
    \end{tikzpicture}
    \hspace{-2mm}}
\foreach \x in {A, ..., Z}
{\expandafter\xdef\csname orcid\x\endcsname{\noexpand\href{https://orcid.org/\csname orcidauthor\x\endcsname}{\noexpand\orcidicon}}}

\begin{document}


\title{ Negative refraction index manipulated by a displaced squeezed Fock state in the mesoscopic dissipative left-handed transmission line  }



\author{Hong-Wei Guo }
\affiliation{Department of Physics, Faculty of Science, Kunming University of Science and Technology, Kunming, 650500, PR China}

\author{ShunCai Zhao\hspace{-1.5mm}\orcidA{}}
\email[Corresponding author: ]{zhaosc@kmust.edu.cn.}
\affiliation{Department of Physics, Faculty of Science, Kunming University of Science and Technology, Kunming, 650500, PR China}

\author{Xiao-Jing Wei }
\affiliation{Department of Physics, Faculty of Science, Kunming University of Science and Technology, Kunming, 650500, PR China}


\begin{abstract}
Negative refractive index (NRI) of the mescopic dissipative left-handed transmission line (LHTL) is manipulated by the displaced squeezed Fock state (DSFS) and the dissipation presented by the resistance and conductance. Comparing to the classical LHTL, some specific quantum characteristics are shown in the LHTL because of quantum effect, which will be significant to its miniaturization application in microwave frequency.
\end{abstract}



\maketitle

\section{Introduction}

The earliest concept of the Left-handed materials was theoretically considered by Veselago \cite{1} in 1968, in which the electric field \(\vec{E}\), magnetic field \(\vec{H}\) and the wavevector \(\vec{k}\) form a left-handed system, which is opposite to the ordinary right-handed media. Many special physical phenomena such as inverse Doppler effects\cite{2,3,4}, reverse Cerenkov radiation\cite{5}, perfect lens \cite{6,7,8,9} occur because of the conuterintuitive physical properties of NRI\cite{10}. Negative-index metamaterial (NIM) transmission lines, also called LHTL existing at microwave frequency band\cite{11,12} are special microwave transmission lines that have been utilized in many applications\cite{13,14,15,16}. Instead of the conventional equivalent circuit model with series inductance and shunt capacitance distributed, a LHTL has series capacitance and shunt inductance distributed which leads to opposite signs of phase and group velocities. For a realistic LHTL, loss occuring in series resistance and shunt conductance is needed in the equivalent circuit model, as shown in Fig.1(a),(b).

There is increasing interest in the nanoelectronics, when the devices or circuits are so small that the transport dimension approaches the charge-carrier inelastic coherent length and the charge carrier confinement dimension reaching the Fermi wavelength, the quantum effects of the devices and circuits must be taken into account\cite{17,18}. With the miniaturization applications of LHTL in mind, we propose a promising new feature in our previous work\cite{19}, in which the quantum influences on NRI of the ideal LHTL are discussed in the thermal Fock state. Here, we investigate the mesoscopic dissipative LHTL in a displaced squeezed Fock state\cite{20,21} which has wide applications in the field of quantum optics\cite{22}, nonlinear problems\cite{23,24,25} and electromagnetic wave\cite{26}. Some novel characteristics of NRI will be shown and be significant for its miniaturization applications in microwave transmission lines.

\section{ The quantum mesoscopic dissipative LHTL}

\begin{figure}[htp]
\center
\includegraphics[totalheight=1.2 in]{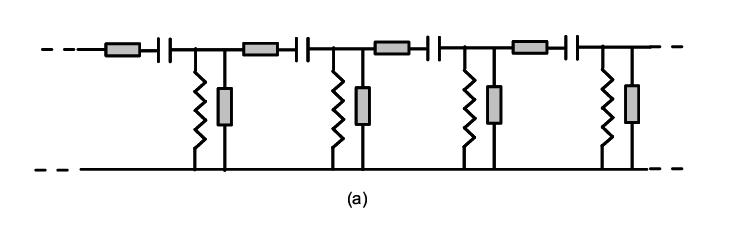 }
\includegraphics[totalheight=1.8 in]{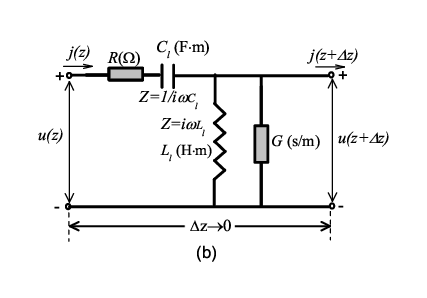 }
\caption{(a) The equivalent circuit model of dissipative left-handed transmission line,
(b) The equivalent unit cell model of dissipative left-handed transmission line}
\end{figure}\label{Fig.1}

We consider LHTL equivalent circuit composed as in Fig.1, which is distributed by series capacitance and shunt inductance, and the present R and G (resistance and conductance) represent dissipation. The classical difference equations , read
\begin{align}
u(z,t)=j(z,t)[R+\frac{1}{i\omega{C_l}}]\Delta{z}+u(z+\Delta{z},t),\\
j(z,t)=u(z,t)[G+\frac{1}{i\omega{L_l}}]\Delta{z}+j(z+\Delta{z},t)
\end{align}
Here \(L_l\), \(C_l\) are the inductance, capacitance of the unit length dissipative LHTL, respectively. \(u(z,t)\) is the voltage, \(j(z,t)\) is the current, and \(\omega\) is the angle frequency.  Considering the seperated current between the inductance \(L_l\) and conductivity \( G\), and the relation between the voltage of the inductance, the solutions to the classical equations of motion corresponding to Eq.(1) and (2) as a consequence of Kirchoff¡¯s law are as follows,
\begin{align}
j(z,t)&= e^{-\sigma z}[A e^{(i\beta{z}-i\omega t)}+A^* e^{(i\omega t-i\beta{z})}],\\
u(z,t)&=\frac{i\omega L_l}{\omega L_l G + 1} e^{-\sigma z}[A^* e^{(i\omega t-i\beta{z})}-A e^{(i\beta{z}-i\omega t)}]
\end{align}
Where \(\sigma\), \(\beta\) are the displacement damping coefficient and wave number, respectively. And their expressions are as follows,
\begin{align}
\sigma&=[\frac{1}{2}[(x-y)+[(x-y)^2+(\frac{G}{\omega C_l}+\frac{R}{\omega L_l})^2]^{\frac{1}{2}}]]^{\frac{1}{2}},\\
\beta&=[\frac{1}{2}[-(x-y)+[(x-y)^2+(\frac{G}{\omega C_l}+\frac{R}{\omega L_l})^2]^{\frac{1}{2}}]]^{\frac{1}{2}}
\end{align}
in which \(x= RG\) and \(y= \frac{1}{\omega^2 L_l C_l}\).

According to the quantization scheme of Louisell\cite{27}, we introduce two variables \(\eta\) and \(\nu\). And they are defined as
\begin{align*}
j(z,t)&=e^{-\sigma z}\eta(z,t),\\
u(z,t)&=\frac{\omega^2}{z_0} e^{-\sigma z}\nu(z,t)
\end{align*}
where \(z_0 \) is the length of per unit cell circuit. It's easy to verify the following relation,
\begin{align}
\frac{\partial}{\partial{\eta}}(\frac{\partial{\eta}}{\partial{t}})+\frac{\partial}{\partial{\nu}}(\frac{\partial{\nu}}{\partial{t}})=0
\end{align}
So \(\eta\) and \(\nu\) are the normal variables. As a result, the Hamiltonian of equivalent unit cell model
of dissipative LHTL is
\begin{align}
\textit{H}=\frac{\omega^2 \nu^2 (\omega L_l G + 1)}{2L_l z_0}+\frac{L_l z_0}{2(\omega L_l G + 1)}\eta^2
\end{align}
We assume that \(\hat{\eta}\) and \(\hat{\nu}\) satisfy the following commutation relation
\begin{align}
[\hat{\eta},\hat{\nu}]=i\hbar=\frac{2i L_l z_0}{\omega_l(\omega L_l G + 1)}[\hat{A},\hat{A^*}]
\end{align}
Then we obtain the quantization of the mesoscopic dissipative LHTL with
\begin{align}
[\hat{A},\hat{A^*}]=\frac{\hbar\omega}{2 L_l z_0}
\end{align}
Where \(\hat{A}\), \(\hat{A^*}\) satisfy
\begin{align}
\hat{A}&=\hat{a}\sqrt{\frac{\hbar\omega (\omega L_l G + 1)}{2 L_l z_0}}, \\
\hat{A^*}&=\hat{a^*}\sqrt{\frac{\hbar\omega (\omega L_l G + 1)}{2 L_l z_0}},
\end{align}
and \([\hat{a}, \hat{a^*}]=1\). So \(\hat{a}\) and \(\hat{a^*}\) are the creation and annihilation operators. \(\hat{a^*}\) should be written as \(\hat{a^\dag}\). The corresponding operators for \(\eta(z,t)\) and \(\nu(z,t)\) can be written as
\begin{align}
\hat{\eta}(z,t)&=\sqrt{\frac{\hbar\omega (\omega L_l G + 1)}{2 L_l z_0}}[\hat{a} e^{(i\beta{z}-i\omega t)}+\hat{a^\dag} e^{(i\omega t-i\beta{z})}],\\
\hat{\nu}(z,t)&=i\sqrt{\frac{ \hbar L_l z_0}{2\omega (\omega L_l G + 1)}}[\hat{a^\dag} e^{(i\omega t-i\beta{z})}-\hat{a} e^{(i\beta{z}-i\omega t)}]
\end{align}
\section{The derivation of NRI in displaced squeezed Fock state}

The displaced squeezed Fock state can be defined as\cite{28,29,30}
\begin{align}
|z, \xi,n\rangle=\hat{D}(z)\hat{S}(\xi)|n\rangle
\end{align}
where the operators \(\hat{D}\) and \(\hat{S}\) satisfy \(\hat{D}(z)= e^(z\hat{a}^{\dag}-z^{*}\hat{a})\), \(\hat{S}(\xi)= e^(\frac{1}{2}\xi \hat{a^\dag}^{2}-\frac{1}{2}\xi^{*}\hat{a^{2}})\), in which $z=|z| e^{i\theta}(|z|>0, 0\leq\theta<2\pi)$  is a displaced parameter, $\xi=|\xi| e^{i\phi}(|\xi|>0, 0\leq\phi<2\pi)$ is a squeezed parameter with the squeezed direction \(\phi\). Using the formula
\begin{align}
e^{\lambda \hat{A}}\hat{B} e^{-\lambda \hat{A}}=\hat{B}+\lambda[\hat{A}, \hat{B}]+\frac{\lambda^{2}}{2}[\hat{A}, [\hat{A}, \hat{B}]]+\cdots
\end{align}
We can achieve the following transformation
\begin{align}
\hat{D^{\dag}}(z)\hat{a}\hat{D}(z)&=\hat{a}+z ,\\
\hat{D^{\dag}}(z)\hat{a^{\dag}}\hat{D}(z)&=\hat{a^{\dag}}+z^{*}
\end{align}
\begin{align}
\hat{S^{\dag}}(\xi)\hat{a}\hat{S}(\xi)&=\hat{a}\cosh|\xi|+\hat{a^{\dag}}e^{i\phi}\sinh|\xi|,\\
\hat{S^{\dag}}(\xi)\hat{a^{\dag}}\hat{S}(\xi)&=\hat{a^{\dag}}\cosh|\xi|+\hat{a}e^{-i\phi}\sinh|\xi|
\end{align}
Therefore, the mean value of the current operator in Heisenberg picture in DSFS can be obtained,
\begin{align}
\langle \hat{j}(z)\rangle &=\langle n,\xi,z|\hat{j}(z)|z,\xi,n\rangle \\ \nonumber
                       &=F e^{-\sigma z}[|z|e^{i\theta}e^{i\beta z}+|z|e^{-i\theta}e^{-i\beta z}]
\end{align}
in which $ F=\sqrt{\frac{\hbar\omega (\omega LG + 1)}{2Lz_{0}}}$, and its corresponding average of the current squared
can be written as
\begin{align}
\langle \hat{j}(z)^{2}\rangle &=\langle n,\xi,z|\hat{j}(z)^{2}|z,\xi,n\rangle    \\ \nonumber
&=F^{2}e^{-2\sigma z}\{[(2n+1)\cosh|\xi|\sinh|\xi|e^{i\phi}+z^{2}]e^{2i \beta z}  \\  \nonumber
&+[(2n+1)\cosh|\xi|\sinh|\xi|e^{-i\phi}+{z^*}^{2}]e^{-2i \beta z }  \\  \nonumber
&+2(n+1)\cosh^{2}|\xi|+2n\sinh^2|\xi|+2z z^*\}
\end{align}
Then the quantum fluctuation of the current should be
\begin{align}
\langle(\Delta \hat{j})^{2}\rangle &=\langle \hat{j}(z)^{2}\rangle -\langle \hat{j}(z)\rangle^{2}  \\  \nonumber
 &=(2n+1)F^{2} e^{-2\sigma z}[\frac{1}{2}\sinh|2\xi|[e^{i\phi}e^{2i\beta z}   \\  \nonumber
 &+e^{-i\phi}e^{-2i \beta z }]+\cosh|2\xi|+1]
\end{align}
It's notice that \(\beta z\) is infinitesimal when z is a dimensionless, using the following transformation relations\cite{11,12,13}
\begin{equation}
\begin{split}
\beta=\omega\sqrt{\mu\epsilon}
\qquad
n_{r}=\frac{c_0 \beta}{\omega} \,
\end{split}
\end{equation}
and Eq.(23), we can obtain the refraction index of the mesoscopic dissipative LHTL with respect to the fluctuation of the current as follows,
\begin{align}
n_{r}=&-\frac{ c_0}{\omega z\sin(\phi)}[\frac{\langle(\Delta \hat{j})^{2}\rangle\frac{2Lz_{0}}{\hbar\omega_{l}(\omega LG + 1)}e^{2\sigma z}-1}{(2n+1)\sinh|2\xi|} \nonumber \\
&-\frac{\cosh|2\xi|}{\sinh|2\xi|}-\frac{\cos(\phi)}{2}]
\end{align}
As mentioned before, its value is negative within the microwave frequency band\cite{11,12,13}.
\section{Result and discussion}

In our model, the resistance \(R\) and conductance \(G\) represent the dissipation in mesoscopic dissipative LHTL, and we set
the relation \(G=\frac{1}{R}\times 10^{-2}\) in Fig. 2. NRI dependent the resistance is controlled by the squeezed parameter in Fig.2. In general, the resistance plays a destructive role in electricity because of the Ohm Law. However, a different feature of \(R\) in the  messoscopic dissipative LHTL appears in the interval of [0, 0.4\(\Omega\)]. It shows that NRI increases when the resistance \(R\) has a fainter increase in the gray block of [0, 0.4\(\Omega\)]. While in the other interval of [0.5\(\Omega\), 2\(\Omega\)], NRI almost remains an invariable value with the same squeezed parameter. At the same time, the effect of the squeezed parameter on NRI is prominent. We notice that the squeezed parameters increase by 0.05\(\pi\), and NRI decreases sharply from \(-1.9\) to \(-0.8\). Comparing to the classic LHTL, the counterintuitive characteristic of the resistance only takes place in the mesoscopic dimension of LHTL.

\begin{figure}[htp]
\center
\includegraphics[totalheight=1.8 in]{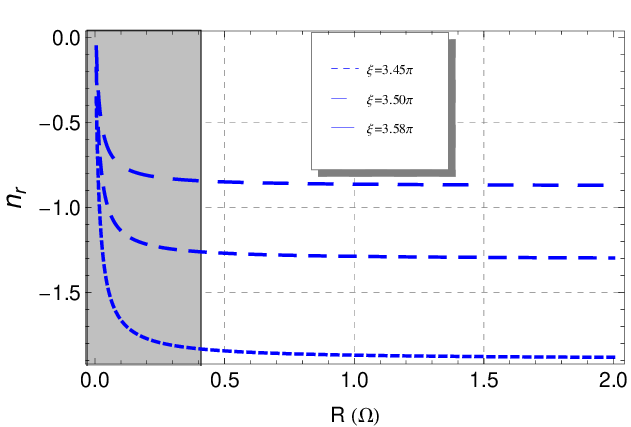 }
\caption{(Color online) The dependence of the negative refraction index on the resistance R of unit cell circuit  with different squeezed parameters. Other parameters are $\omega=3 GHz$, \(\langle(\Delta \hat{j})^{2}\rangle=10\),\(\phi=\frac{\pi}{3}\) \(n=2\), \(z_0=4z=4\mu m\), $L=398 \mu H$, $C=995 p F$.}
\end{figure}\label{Fig.2}

Another interesting regulatory parameters are the squeezed parameters. Fig.3 shows the influence of squeezed direction on NRI in DSFS, i.e., NRI versus the squeezed direction of [0,\(\pi\)] with different photon numbers 1, 2, 15. The two gray blocks correspond to two angular intervals [0, 0.2\(\pi\)] and [0.8\(\pi\), 1.0\(\pi\)], respectively, which means two opposite squeezed directions. The interval of [0, 0.2\(\pi\)] shows the squeezed direction along transmission line, and an inverse squeezed direction happens in the interval of [0.8\(\pi\), 1.0\(\pi\)]. In the forward squeezed direction, NRI decreases when the direction angular enhances, while NRI increases in the reverse squeezed direction. Utilizing Eq. (8), (13) and (14), the quantum Hamiltonian of the dissipative LHTL reaches to \(\hat{\textit{H}}=\hbar \omega (\hat{a^\dag}\hat{a}+\frac{1}{2})\). Therefore, the energy of the electromagnetic wave in LHTL is proportional to the photon numbers. Fig.3 shows the amount of the energy of the electromagnetic wave reduces NRI, which is a novel quantum behavior in the mesoscopic LHTL.

\begin{figure}[htp]
\center
\includegraphics[totalheight=1.8 in]{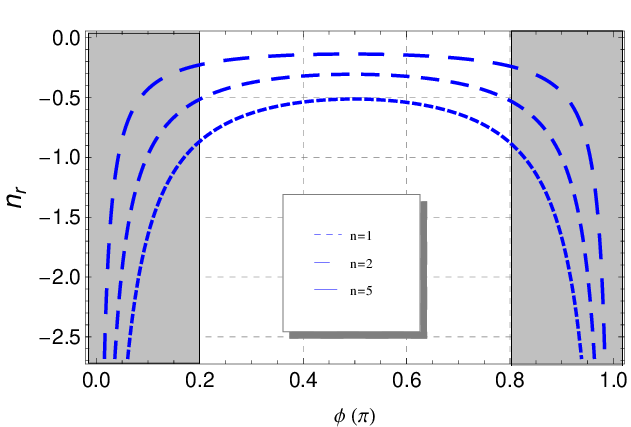 }
\caption{(Color online) The dependence of the negative refraction index on the squeezed direction $\phi$ with different photon numbers \(n= \)1, 2, 15. $\xi=2.8\pi$, $G=0.2S/m$, $R=0.2 \Omega$, and Other parameters are same to Fig.2.}
\end{figure}\label{Fig.3}

In the classic LHTL, NRI is a constant alone the circuit when the circuit is in the steady status. However, something may be different in the mesoscopic LHTL. Fig. 4 plots NRI alone the circuit within a unit cell \(z_0\), and NRI shows some attractive features because of the quantum effect. We note that NRI decrease alone the circuit within a unit cell \(z_0\), and it reaches asymptotically to a constant in spite of the different squeezed parameters. The larger squeezed parameter leads to a smaller NRI. That NRI alone the circuit within the unit cell has an gradient value is distinctive to the LHTL in the macro scale. This feature may be significant and interesting to miniature design and practical applications for LHTL.

\begin{figure}[htp]
\center
\includegraphics[totalheight=1.8 in]{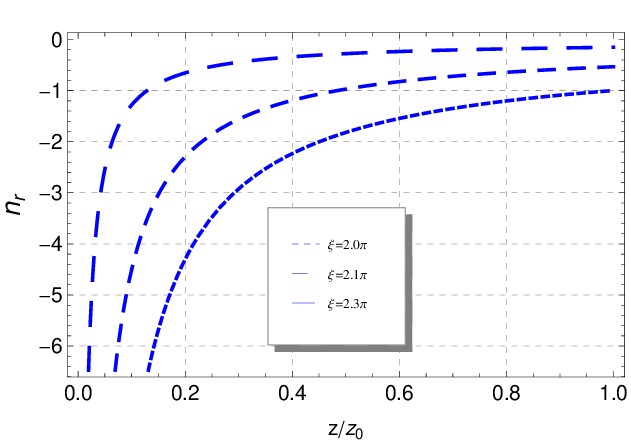 }
\caption{(Color online) The dependence of the negative refraction index on the displacement squeezed parameters with different squeezed parameters. \(n=5\), $G=0.02 S/m$, $R=0.02\Omega$ \(\phi=\frac{\pi}{5}\), and other parameters are same as Fig.3.}
\end{figure}\label{Fig.4}

\section{conclusion}

The quantum characteristic of NRI was manipulated by the DSFS and dissipation in the mesoscopic dissipative LHTL. Alone the transmission line, NRI is not homogeneous but decreases to an asymptotic value with different squeezed parameters, and the forward and inverse squeezed direction results in the decreasing and increasing NRI alone the transmission line. The tiny dissipation, i.e., the resistance or conductance leads to the increasing NRI, while the increasing resistance brings out an asymptotic constant of NRI. The effect of DSFS and dissipation caused by quantum effect will be significant to NRI in its miniaturization applications in the future.

\begin{acknowledgments}
This work is supported by the National Natural Science Foundation of China ( Grant Nos. 61205205 and 6156508508 ), the General Program of Yunnan Applied Basic Research Project, China( Grant No. 2016FB009 ) and the Foundation for Personnel training projects of Yunnan Province, China ( Grant No. KKSY201207068 ).
\end{acknowledgments}






\end{document}